\documentclass[universe,article,authorversion,moreauthors,LaTeX, dvi2pdf,10pt,a4paper,tikz]{mdpi} 

\firstpage{1} 

\pubvolume{1}
\articlenumber{0}
\pubyear{2024}
\copyrightyear{2024}

\usepackage{graphicx}	
\usepackage{amsmath}
\usepackage{amssymb}

\Title{Strange Dwarfs: a review on the (in)stability}

\Author{Francesco Di Clemente $^{1}$\orcidA{}, Alessandro Drago $^{1,2,*}$\orcidB{}, and Giuseppe Pagliara $^{1,2}$\orcidC{}}

\AuthorNames{Francesco Di Clemente, Alessandro Drago, and Giuseppe Pagliara}

\address{$^{1}$INFN, Sezione di Ferrara, via Saragat 1, I-44122 Ferrara, Italy\\
$^{2}$Department of Physics and Earth Science, University of Ferrara, via Saragat 1, I-44122 Ferrara, Italy}

\corres{Correspondence: drago@fe.infn.it}

\abstract{White dwarfs are the remnants of stars not massive enough to become supernovae. This review explores the concept of strange dwarfs, a unique class of white dwarfs which contain cores of strange quark matter. Strange dwarfs have different sizes, masses, and evolutionary paths with respect to white dwarfs. They might form through the accumulation of normal matter on strange quark stars or by capture of strangelets. The stability of strange dwarfs has been debated, with initial studies suggesting stability, while later analyses indicated potential instability. This review revisits these discussions, focusing on the critical role of boundary conditions between nuclear and quark matter in determining stability. It also offers insights into their formation, structure, and possible detection in the universe.}

\keyword{Strange matter; White dwarfs; Accretion induced collapse} 

\begin{document}

\section{\label{sec:introduction}Introduction}
White dwarfs (WDs) are astrophysical objects that originate from the remnants of stars whose initial mass was below approximately $9\,M_{\odot}$ \citep{Heger:2002by}. These stars, after depleting their reserves of nuclear fuel, enter a phase during which their core contracts because the nuclear reactions cannot counteract anymore gravitational force, while their outer layers expand. This collapse is halted only when the electrons within the star core become degenerate, providing the necessary pressure to counteract further gravitational collapse.
Depending on the mass of the progenitor, the stellar outcome can be different. Indeed, the nuclear fusion reactions that occur during the star's evolution can lead to the production of different types of nuclei, ultimately influencing the nature of the resulting WDs. These include helium (He), carbon-oxygen (C-O), and oxygen-neon-magnesium (O-Ne-Mg) WDs. It's essential to note that the maximum mass that a WD can attain, referred to as the Chandrasekhar mass and calculated to be approximately $1.4\,M_{\odot}$ \citep{Chandrasekhar:1931ih}, varies depending on the composition of the WD. In practice, the majority of observed WDs are of the C-O type.

In \citet{Glendenning:1994sp,Glendenning:1994zb} it was proposed that WDs could possess an inner core composed of absolutely stable strange quark matter. This is a consequence of the Bodmer-Witten hypothesis \citep{Bodmer:1971we,Witten:1984rs}. What makes this idea even more interesting is that the presence of this stable strange quark matter core has the potential to make some of these compact objects stable, while the corresponding configuration without the strange quark core which would be unstable.

These objects, named "strange dwarfs" (SDs), exhibit characteristics distinct from those of conventional WDs. Specifically, SDs can possess radii, masses, and astrophysical evolution pathways different from those of standard WDs. It was conjectured that SDs could form either by accumulating normal nuclear matter on the surface of a strange quark star (SQS) or by collecting clusters of strange quark matter, commonly referred to as "strangelets", onto WDs.

\citet{Glendenning:1994sp} studied the radial stability of SDs, showing that these objects can remain stable even if the density of the nuclear matter envelope surpasses the maximum density observed in typical WDs.

The question concerning the stability of SDs underwent a thorough reexamination in the work of \citet{Alford:2017vca}. They found the eigenvalue associated with the fundamental radial mode to be negative, indicating that SDs, in fact, are unstable.

It was suggested that the previous works by \citet{Glendenning:1994sp,Glendenning:1994zb} may have inadvertently misinterpreted their findings by confusing the second-lowest eigenmode with the lowest one. However, upon closer examination and analysis, it became evident that these two sets of research were built upon different underlying hypotheses \citep{DiClemente:2022ktz}.
Contrary to the initial belief that the two studies were grounded on the same assumptions, it became clear that they actually operated within slightly distinct theoretical frameworks, each with its own validity. This realization effectively solved the apparent contradiction between the results obtained by the two works.
Specifically, \citet{DiClemente:2022ktz} focused on the boundary conditions at the interface where nuclear matter meets the quark core within SDs.
The analysis is based on part of the formalism established in previous works, in particular \citet{Pereira:2017rmp} and \citet{DiClemente:2020szl}. These studies provide insights into the boundary conditions that ought to be applied in the context of rapid (and slow) conversions between nuclear matter and quark matter and in the context of phase transitions in general. Crucially, the specific boundary conditions employed can exert a substantial influence on the eigenvalues governing radial oscillations, and, by extension, they can have a profound impact on the overall stability of the star.
In addition, \citet{DiClemente:2022ktz} also addresses the applicability of the traditional stability criterion based on the analisys of the extrema of the trajectory in the MR plane \citep{zeldovich:1963,bardeen:1966}. However, in this specific case, a crucial refinement is added to this criterion, by emphasizing the need for explicit specification regarding whether the quark content of the star remains constant or undergoes changes during the radial oscillations.

In this review, we build upon prior research by incorporating a comprehensive analysis of the equations of state (EoSs) relevant to SDs. This involves a detailed examination of the mass-radius relationship for these objects. Furthermore, we give an explicit mathematical framework that yields a formula for fixing the quark content within the core of an SD.

Additionally, we have broadened our discussion to include potential astrophysical signatures of SDs. This extension is crucial for observational astrophysics, as it may provide insights into identifying and verifying the existence of SDs through observable phenomena. 

\section{Equation of state}

One of the important considerations in assessing the stability of SDs lies in the nature of their EoS. Historically, when examining this aspect in previous works \citep{Glendenning:1994sp,Glendenning:1994zb,Alford:2017vca}, the EoS was formulated as:

\begin{equation}
\varepsilon(P)=\begin{cases}
\varepsilon_\mathrm{BPS}(P) & \text{if}\ P \leq P_\mathrm{t}\\
\varepsilon_\mathrm{quark}(P) & \text{if}\ P>P_\mathrm{t} \, .
\end{cases}\label{eq:piecewiseEOS}
\end{equation}
In this expression, $\varepsilon_\mathrm{BPS}$ represents the Baym-Pethick-Sutherland (BPS) EoS \citep{Baym:1971pw}, while $\varepsilon_\mathrm{quark}$ denotes an EoS characterizing strange quark matter. An example of such an EoS could be the one based on the MIT bag model \citep{Chodos1974}. The critical parameter in this formulation is the transition pressure, denoted $P_\mathrm{t}$, which is defined as the pressure at the interface between quarks and nuclear matter. 

As EoS for nuclear matter we use the BPS EoS which represents the ideal "limit" WDs in which all elements up to Fe have been produced. This EoS displays a Chandrasekhar mass of approximately $1 M_\odot$, a value lower than the typical Chandrasekhar mass for C-O WDs or O-Mg-Ne WDs ($1.4 M_\odot$ and $1.2 M_\odot$). 
It has been pointed out in \citet{Benevnuto:1996} that the use of the BPS EoS is not realistic for WDs. Nevertheless, as it will be explained in \autoref{subsec:kurban}, it represents the limit in compactness for WDs, since it provides the smallest radius for a given mass. 

In our work we use a fit of the BPS EoS in order to avoid artifacts due to the numerical differentiation of a piecewise interpolation. The form of the fit of the BPS EoS reads:

\begin{equation*}
    \varepsilon(P)=e^{f(ln(P))}
\end{equation*}

where the function $f$ is

\begin{align*}
  f(x)=&-1496.7952111882255 + 1109.8179718329682\, x^{1/3}\\& - 
 171.06847907037277\, x^{4/3} + 95.47548413371702\, x^{5/3}\\& - 
 19.652076548674618\, x^2 + 1.4412357260222872\, x^{7/3}\\& - 
 3.995517504571193\times 10^{-14} x^8.
\end{align*}

\begin{figure}[t]
\begin{centering}
\includegraphics[width=0.7\textwidth]{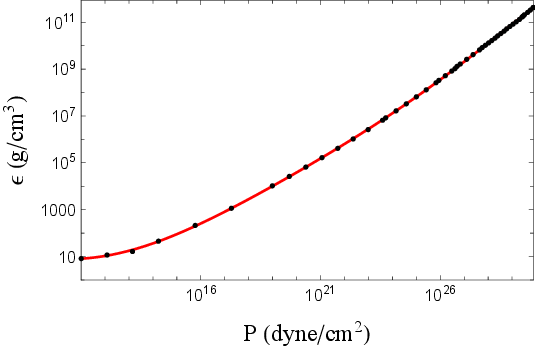}
    \caption{BPS equation fit in red, tabulated points in black}
    \label{bpsfit}
    \end{centering}
    
\end{figure}

Note that the inclusion of decimal digits is essential for achieving adequate precision across a broad spectrum of pressures and densities. The fit ranges in energy density from $\sim 7$g/cm$^3$ to $\sim 4\times 10^{11}$g/cm$^3$ and it is visible in  \autoref{bpsfit}.

When dealing with SDs, it's important to understand that one can use any value for $\varepsilon_\mathrm{t}$ as long as it's less than $\varepsilon_\mathrm{drip} \approx 4\times 10^{11}$ g/cm$^3$.

Unlike regular WDs, where one typically only needs to specify the central pressure $P_0$ to define a star's configuration when solving the TOV equation \citep{Oppenheimer:1939ne}, SDs require two parameters. As clear from  \autoref{eq:piecewiseEOS}, the first parameter is the transition pressure  $P_\mathrm{t}$, which represents the pressure at the interface where the quark core meets the outer nuclear matter envelope, the second one is indeed the central pressure as in the normal case  $P_0$.

What allows the formation of SDs is the existence of the Coulomb barrier that separates the outer nuclear matter from the inner core of quark matter. This separation occurs under the condition that the maximum density of nuclear matter remains lower than the neutron drip density $\varepsilon_\mathrm{drip}$. Beyond this density, free neutrons start to appear. Importantly, since they are not subject to the constraints of the Coulomb barrier, they can readily penetrate the core of quark matter. Upon entering the core, they are absorbed, leading to the deconfinement of their constituent quarks.

Given that the solutions of the TOV equation for SDs depend on two parameters, a question arises about the suitability of choosing the pair of parameters $(P_0, P_\mathrm{t})$ for characterizing these configurations. Choosing a value of $P_\mathrm{t}$ does not account for the fact that below the neutron drip density the baryonic content of the core remains constant, despite changing the central pressure $P_0$.  The studies by \citet{Vartanyan:2009zza} and \citet{Vartanyan:2012zz} discuss the case in which nuclear matter cannot transition into quark matter, allowing for the definition of sequences of configurations that have the same quark baryon number in the core, that we define as $B_{\mathrm{core}}$. Consequently, one can solve the TOV equation with alternative parameter pair, namely $(P_0, B_{\mathrm{core}})$.

The quark baryon number, represented as $B_\mathrm{core}$, can be expressed as follows:

\begin{equation}
B_\mathrm{core}(P_0,P_\mathrm{t}) = \int_{0}^{R_\mathrm{core}} {4 \pi r^2 \frac{\rho(r)}{\sqrt{1-2 m(r)/r}}\,dr}. \label{bcore}
\end{equation}

Here, $\rho$ is the baryon density within the quark core. It's important to note that these two parameter choices are not interchangeable. If one opts to keep $P_t$ constant while varying $P_0$, this leads to changes in $B_{\mathrm{core}}$, implicitly indicating a scenario in which hadrons can deconfine into quarks, because the change in the core is encoded in the fact that the central pressure is changing by fixing the external pressure of the core $P_t$ (transition pressure). Conversely, when $B_{\mathrm{core}}$ is maintained at a constant value, one necessitates to increase in $P_t$ with higher values of $P_0$, illustrating a situation in which hadrons accumulate on the surface of the strange core without undergoing transformation into quarks.

In order to correctly consider a SD EoS at its equilibrium, we want to chose the parameter pair $(P_0, B_{\mathrm{core}})$. $B_{\mathrm{core}}$ is a function of $P_0$ and $P_\mathrm{t}$, therefore, we need to find the inverse relation that, given a choice of $B_{\mathrm{core}}$, gives back the value of $P_\mathrm{t}$.

In our analysis, since the core of the system is relatively small, we can reasonably approximate the behavior using Newtonian physics in a first-order approximation.

For the EoS governing a small quantity of quark matter within the core, we utilize a parametric expression:

\begin{equation}
P=(\varepsilon-\varepsilon_\mathrm{W})a.\label{eq:naivequark}
\end{equation}

Here, the variable $a$ serves as a multiplicative parameter that encompasses various factors, including the bag constant. The term $\varepsilon_\mathrm{W}$ corresponds to the Witten density, defined as $\varepsilon_\mathrm{W}=\varepsilon(P=0)$. From considering the Newtonian hydrostatic equilibrium and \autoref{eq:naivequark}, we obtain:

\begin{equation}
a \frac{d\varepsilon}{\varepsilon^2}=-\frac{4 \pi}{3} r\,dr.
\end{equation}

Upon integrating both sides and combining all constants into a single parameter, denoted as $K$, we obtain the following equation:

\begin{equation}
\int_{\varepsilon_0}^{\varepsilon_\mathrm{W}} \frac{d\varepsilon}{\varepsilon^2}=
-K \int_0 ^{R_\mathrm{core}} dr\,r .
\end{equation}

In this equation, $R_\mathrm{core}$ represents the radius of the core and $\varepsilon_0$ the energy density at the center of the system. The solution to both sides of this equation leads us to the following expression:

\begin{equation}
\varepsilon_0 = \frac{\varepsilon_\mathrm{W}}{1- K \varepsilon_\mathrm{W} R_\mathrm{core}^2} .\label{eq:fixedcoreW}
\end{equation}

This equation shows how the central energy density is connected to the Witten density $\varepsilon_\mathrm{W}$, considering the parameter $K$ and the core radius $R_\mathrm{core}$.

We can modify  \autoref{eq:fixedcoreW} to replace the core radius $R_\mathrm{core}$ with its baryon content, denoted as $B_{\mathrm{core}}$, since $R_\mathrm{core}^2 \propto B_{\mathrm{core}}^{2/3}$. Additionally, considering that our core experiences slight compression due to the surrounding nuclear matter, we can replace $\varepsilon_\mathrm{W}$ with the effective energy density specific to the quark core, denoted as $\varepsilon_\mathrm{t}^\mathrm{Q}$, which is slightly greater than $\varepsilon_\mathrm{W}$. Indeed, the transition density at the boundary of the quark core satisfies $\varepsilon_\mathrm{t}^\mathrm{Q}\geq \varepsilon_\mathrm{W}$. The new form of the equation reads:

\begin{equation}
\varepsilon_0 (B_{\mathrm{core}}) = \frac{\varepsilon_\mathrm{t}^\mathrm{Q}}{1- K \varepsilon_\mathrm{t}^\mathrm{Q}B_{\mathrm{core}}^{2/3}}. \label{eq:fixedcore1}
\end{equation}

Now, it's reasonable to incorporate some higher-order corrections into \autoref{eq:fixedcore1} to account for general relativistic effects:

\begin{equation}
\varepsilon_0 (B_{\mathrm{core}}) = \frac{\varepsilon_\mathrm{t}^\mathrm{Q}}{1- k_1 \varepsilon_\mathrm{t}^\mathrm{Q}B_{\mathrm{core}}^{2/3}- k_2 \varepsilon_\mathrm{t}^\mathrm{Q}B_{\mathrm{core}}^{4/3}}, \label{eq:fixedcore2}
\end{equation}

In this modified equation, we introduce two numerical parameters, $k_1$ and $k_2$, to account for these higher-order relativistic effects. These parameters are determined numerically and they depend on the quark matter EoS.

For the quark matter, we utilize the following thermodynamic potential:

\begin{equation}
\Omega (\mu)= -\frac{3}{4 \pi^2} a_4 , \mu^4 +\frac{3}{4\pi^2}(m_s^2 -4 \Delta_0^2)\mu^2+B.\label{eq:eosquarki}
\end{equation}

Here $a_4=0.7$, the gap parameter $\Delta_0$ = 80 MeV, the strange quark mass $m_s$=120 MeV, and the bag constant $B=135^4 $ MeV$^4$, as in \citet{Bombaci:2020vgw}.

It's important to note that the parameters $k_1$ and $k_2$ are primarily influenced by the bag constant, and any reasonable adjustments made to the other parameters in \autoref{eq:eosquarki} have negligible impacts on these parameters.

\subsection{Mass-radius relation}
The relationship between mass and radius exhibits notable differences depending on whether we consider the parameter pairs $(P_0,P_\mathrm{t})$ or $(P_0,B_{\mathrm{core}})$. When constructing a mass-radius (MR) diagram, it's essential to vary one parameter, typically the central pressure (or central energy density), while keeping the other parameters constant.

\begin{figure}
\begin{centering}
\includegraphics[width=0.7\textwidth]{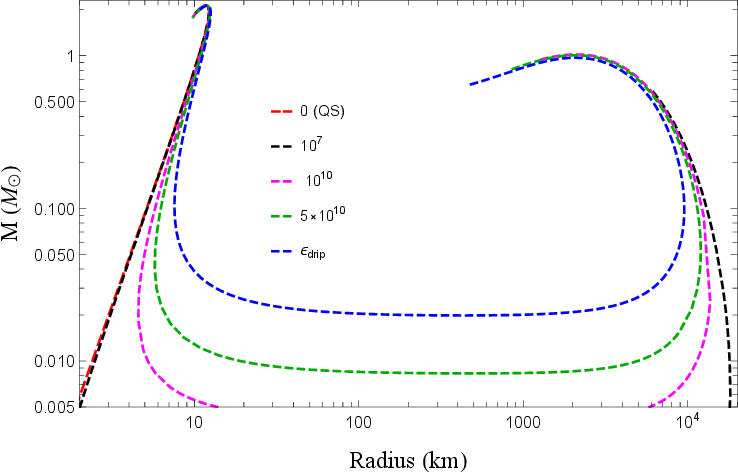}
    \caption{MR diagram for EoSs having fixed transition densities, indicated in the legend in units of g/cm$^3$. In dashed red is shown also the curve for a bare SQS, which indeed does not have a transition density to nuclear matter.}
    \label{MRdiagram1sd}
\end{centering}
\end{figure}

\begin{figure}
\begin{centering}
\includegraphics[width=0.7\textwidth]{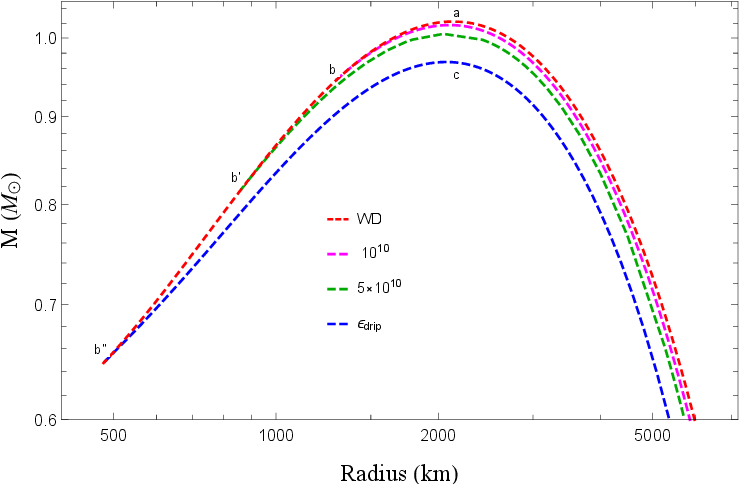}
    \caption{Enlarged view of the MR sequence \citep{DiClemente:2022ktz} approaching the Chandrasekhar limit, using the same notation as in \autoref{MRdiagram1sd}. Additionally, the WD configuration is showed for comparison.}
    \label{MRzoom}
\end{centering}
\end{figure}

If we opt to fix the transition pressure $P_\mathrm{t}$, we are essentially exploring configurations where quark matter consistently appears at the same energy density threshold. We begin with a configuration where $P_0=P_\mathrm{t}$, denoted as point $b$ (or $b'$ and $b''$ depending on the transition energy density), as illustrated in  \autoref{MRzoom}. Then, we progress along the bottom branch in a clockwise direction, increasing the central pressure.

In  \autoref{MRzoom}, also it becomes clear how the extreme point corresponding to the BPS EoS, indicated as "WD", joins with the curve obtained by fixing $\varepsilon_\mathrm{t} = \varepsilon_\mathrm{drip}$. Indeed, the curves built by choosing the transition any pressure join exactly at the point in which, for the BPS,  $P_0=P_\mathrm{t}$ (or $\varepsilon_0 = \varepsilon_\mathrm{t}$). When the transition density is relatively low, the point where the curves join falls before reaching the WDs maximum mass on the MR diagram. This is evident in  \autoref{MRdiagram1sd}, where a low energy transition density ($\varepsilon_\mathrm{t}=10^7$ g/cm$^3$) is represented by the dashed black curve, which has its maximum allowed mass at approximately 0.5 $M_\odot$: this is where it joins the WD curve, which isn't shown in the figure.

\begin{figure}
\begin{centering}
\includegraphics[width=0.7\textwidth]{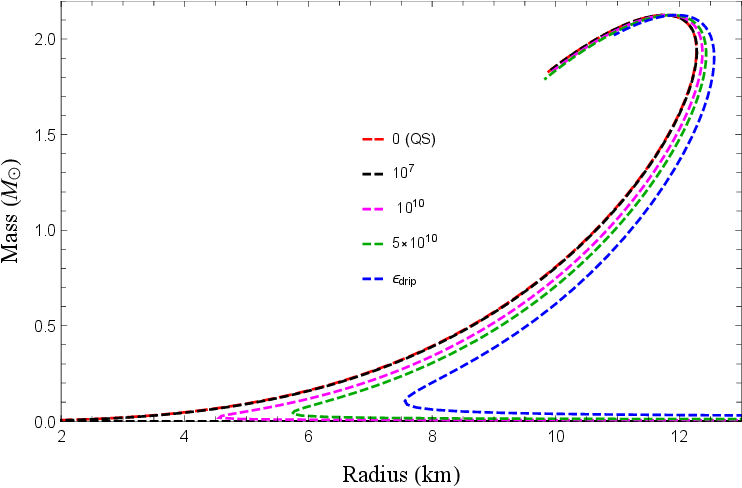}
    \caption{Detail of the \autoref{MRdiagram1sd}. Here it is visible the SQSs branch, where EoSs having a constant $\varepsilon_\mathrm{t}$ are analogous to non-bare SQSs, namely SQSs with a nuclear crust.}
    \label{sddetails1}
\end{centering}
\end{figure}

\begin{figure}[H]
\begin{centering}
\includegraphics[width=0.7\textwidth]{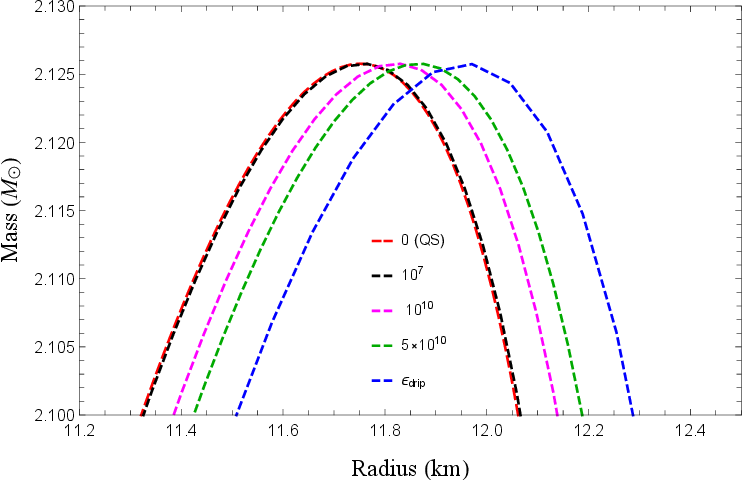}
    \caption{Magnification of the maxima in \autoref{sddetails1}. The higher is the transition density of the nuclear matter and the more compact is the star. Indeed, the wider the range between the constant $\varepsilon_\mathrm{t}$ and the star's surface, the greater the compression exerted by the surrounding nuclear matter on the strange core.}
    \label{sddetails2}
\end{centering}
\end{figure}

On the other hand, the MR relation for SDs tends to converge with that of a SQS when considering small radii. In particular, if the value of $\varepsilon_\mathrm{t}(P_\mathrm{t})$ is significantly smaller than the neutron drip density, it implies that there isn't sufficient matter above the quark core (which now constitutes the majority of the star) to exert significant compression on it. For the high values of $\varepsilon_\mathrm{t}(P_\mathrm{t})$, specifically the neutron drip density, the radii of SDs are slightly smaller compared to those of a SQS. This occurs because there is a broader range of pressure that nuclear matter must cover in these cases. This behavior is showed in the  \autoref{sddetails1}.

\begin{figure}[H]
\begin{centering}
\includegraphics[width=0.7\textwidth]{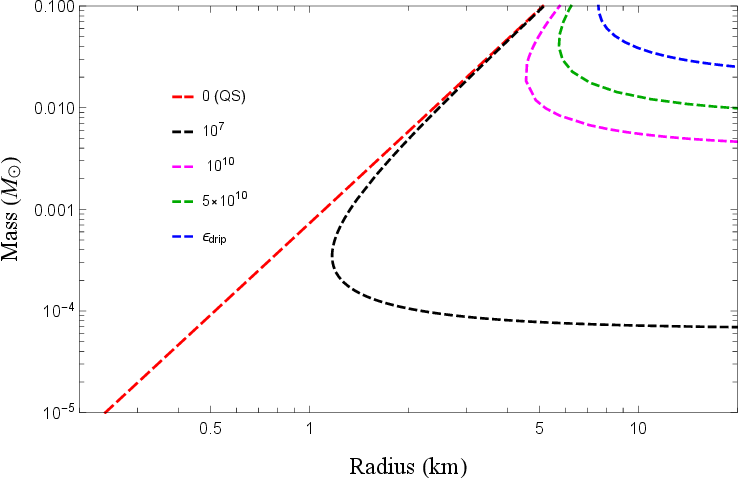}
    \caption{A closer look at the low-mass stars in \autoref{sddetails1}. It is evident that when $\varepsilon_\mathrm{t}$ is low, the deviation from the radius of a SQS is less pronounced.}
    \label{sddetails3}
\end{centering}
\end{figure}

When we fix $\varepsilon_\mathrm{t}(P_\mathrm{t})$, we implicitly let $B_{\mathrm{core}}$ vary. Conversely, when we choose to fix the baryon content of the core ($B_{\mathrm{core}}$), it is the transition density that changes along the diagram. 

When we establish a fixed value for $B_{\mathrm{core}}$, the corresponding configuration contains a specific quantity of quarks in its core. If we start from a point at which $\varepsilon_\mathrm{t}(P_\mathrm{t})=0$ there is no nuclear matter situated above core to exert compression. In other words, it corresponds to the extreme point on the left side of  \autoref{MRdiagram1sd} (red dashed curve) and satisfies \autoref{eq:fixedcoreW}.

As we increase the central energy density (which means adding nuclear matter on top of the quark core), we move in a counter-clockwise direction on the MR diagram. During this progression, we intersect curves in  \autoref{MRdiagram1sd} that correspond to increasing values of $\varepsilon_\mathrm{t}(P_\mathrm{t})$. This means that a curve representing a constant $B_{\mathrm{core}}$ is comprised of configurations with varying $\varepsilon_\mathrm{t}(P_\mathrm{t})$. The initial point on this curve has $\varepsilon_\mathrm{t}=0$ (or equivalently $\varepsilon_\mathrm{t}^\mathrm{Q}=\varepsilon_\mathrm{W}$), while the final point corresponds to $\varepsilon_\mathrm{t}=\varepsilon_\mathrm{drip}$.

\begin{figure}[H]
\begin{centering}
\includegraphics[width=0.7\textwidth]{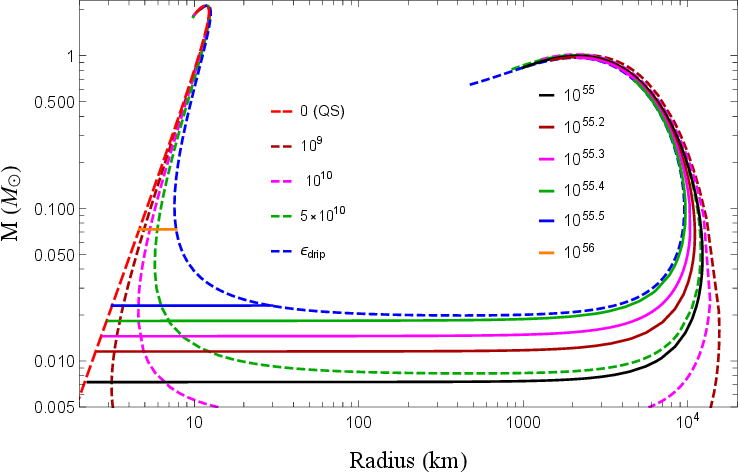}
    \caption{MR sequences of SDs. Configurations with constant transition pressure $P_t$ are represented by dashed lines. As the central pressure $P_0$, namely $B_{\mathrm{core}}$, increases, these sequences progress in a clockwise direction. The corresponding values of the transition density $\varepsilon_t$, measured in $\mathrm{g/cm^3}$, are specified in the legend. Conversely, solid lines depict configurations where the quark content of the core, $B_{\mathrm{core}}$, remains unchanged. In this scenario, increasing $P_0$ (therefore increasing $P_t$) makes the curves progress anti-clockwise. In this case the legend indicates the specific values of $B_{\mathrm{core}}$.
    }
    \label{mrfixedcore}
\end{centering}
\end{figure}

In  \autoref{mrfixedcore}, we can observe a specific behavior where, if the value of $B_{\mathrm{core}}$ is too large, the condition $\varepsilon_\mathrm{t}=\varepsilon_\mathrm{drip}$ is achieved at relatively small radii. The extreme points belong to the curve representing the highest possible density of nuclear matter within the star, which corresponds to the neutron drip density (the blue dashed one).

\section{\label{subsec:radial}Radial oscillations}

To assess the stability of a star, radial oscillations are a valuable analytical tool. The equation for radial oscillations is derived by perturbing both the fluid variables and the spacetime metric that characterizes the interior of the star.

Using this metric, the differential equation governing radial oscillations can be expressed as:

\begin{equation}
(H \xi')'=-(\omega^2 W + Q)\xi,
\label{eq:radial}
\end{equation}

Here, $\xi(r)$ represents the rescaled radial Lagrangian displacement  \citep{bardeen:1966}, while $\omega$ is the characteristic frequency of the oscillation mode. The functions in  \autoref{eq:radial} are defined as:

\begin{align}
H &=  r^{-2} (\varepsilon + P)  e^{\lambda + 3 \phi} c_s^2 \nonumber \\
Q &=  r^{-2} (\varepsilon + P) e^{\lambda + 3 \phi} (\phi'^2 + 4 r^{-1}\phi'-8 \pi e^{2 \lambda}P) \nonumber  \\
W &= r^{-2} (\varepsilon + P)  e^{3\lambda +  \phi}\,,
\label{eq:coefficinets_SL}
\end{align}

Here, $c_s^2$ represents the speed of sound, while $\lambda(r)$ and $\phi(r)$ are the metric potentials. It's essential to note that when dealing with multiple layers or phase transitions within the star, it becomes necessary to establish clear boundary conditions at the interfaces between these layers. This aspect is discussed in \citet{Pereira:2017rmp} and \citet{DiClemente:2020szl}. In particular, one must specify whether, within the timescale of the oscillation, the two components of the fluid can transition into one another. This consideration depends on the presence of phase transitions and their associated timescales. Therefore, we categorize these transitions as either \textit{slow transitions} or \textit{fast transitions} to distinguish between their characteristic timescale.

\subsection{\label{subsec:slow}Slow transition}
The scenario of a slow phase transition holds when the timescale for the conversion from one phase to another is significantly longer than the timescale of the perturbation itself. In this scenario, the two phases do not intermix during the oscillations, and the volume element near the surface that separates the phases moves along with the interface, expanding and contracting. This particular situation is applicable to SDs where $\varepsilon_\mathrm{t} < \varepsilon_\mathrm{drip}$, which essentially means that the two phases never mix. In other words, we are considering the stability of star configurations on the MR diagram where $B_\mathrm{core}$ remains constant, as it cannot increase. In a practical sense, each star configuration can be associated with either a curve characterized by $B_\mathrm{core}$ or one defined by $P_\mathrm{t}$ but only one of them is physically acceptable. Consequently, the choice of boundary conditions for radial oscillations must align with this physical interpretation.
\begin{figure}[t]
\begin{centering}
    \includegraphics[width=0.7\textwidth]{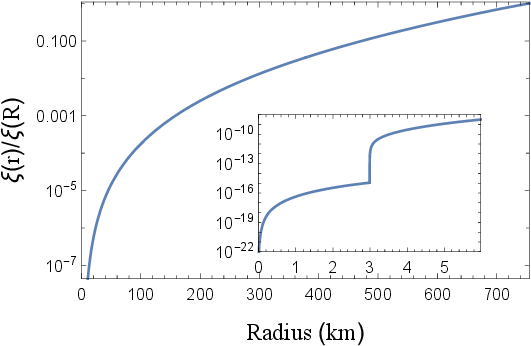}
    \caption{In the scenario where hadronic matter does not transition into quark matter during the oscillation timescale, the fundamental eigenfunction of radial modes is examined. The star under consideration in this analysis and referred to in the fast scenario figure has  M$\,\simeq 0.02 \,\mathrm{M}_\odot$,  $B_\mathrm{core} \simeq 2.69 \times 10^{55}$ and $\varepsilon _t = \varepsilon_\mathrm{drip}$. This strange dwarf is situated just beyond the minimum point on the dashed blue curve in \autoref{mrfixedcore}. The mode in question is stable since $\omega^2 = 0.788275 \mathrm{\,Hz^2}$ is positive. A closer look at the region near $r=r_t$ within the inset plot highlights a kink in the eigenfunction.}
    \label{slow}
    \end{centering}
\end{figure}

For the slow conversion, the interface conditions involve maintaining the continuity of the radial displacement at the boundary radius $r_\mathrm{t}$:

\begin{equation}\label{eq:slow1}
    \left[\xi \right]^+_- \equiv \xi(r_\mathrm{t}^+ )-\xi(r_\mathrm{t}^-)=0\,,
\end{equation}

Additionally, it requires ensuring the continuity of the Lagrangian perturbation of pressure:

\begin{equation}\label{eq:slow2}
    \left[\Delta P\right]^+_- = \left[ -e^\phi\, r^{-2}\, \gamma (r) \, P \, \frac{\partial \xi}{\partial r} \right]^+_-=0 \,,
\end{equation}

Here, $\gamma(r)$ represents the relativistic adiabatic index, given by $\gamma(r)=(\partial P /\partial \varepsilon)(\varepsilon+P)P^{-1}$. By solving \autoref{eq:radial} applying these conditions, we obtain that $\omega^2>0$, and it vanishes at the maximum mass in the MR plane along the curve defined by the constant values of $B_\mathrm{core}$, as proposed by the criterion of \citet{zeldovich:1963} and \citet{bardeen:1966}. The eigenfunctions exhibit continuity with a kink at $r_\mathrm{t}$ (as shown in  \autoref{slow}), and the same behavior is reflected in $\Delta P (r)$ \citep{DiClemente:2020szl}.

\subsection{\label{subsec:fast}Rapid transition}
When the timescale for the conversion between two phases is shorter than the timescale of the perturbation, the exchange of mass between these phases becomes possible. The boundary between these phases is in thermodynamic equilibrium, given the rapid conversion rates, therefore,  \autoref{eq:slow2} remains applicable in this scenario.

The main difference compared to the slow transition case lies in the interface condition from  \autoref{eq:slow1}. In this scenario, it transforms into:

\begin{equation}\label{eq:rapid}
    \left[\xi  + \frac{\gamma P \xi'}{P'}\right]^+_- =0\,,
\end{equation}

\begin{figure}[t]
\begin{centering}
\includegraphics[width=0.7\textwidth]{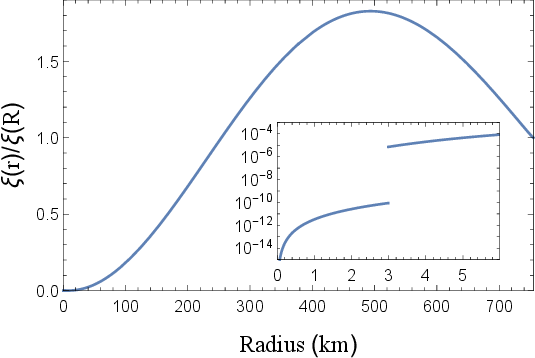}
    \caption{In the rapid transition scenario we observe a fundamental eigenfunction that has has a sharp discontinuity, suggesting an instantaneous change rather than a gradual one, even though extremely rapid as in \citet{Alford:2017vca}. This mode is unstable, since it is characterized by a negative squared frequency of $\omega^2 = -1.62785 \mathrm{\,Hz^2}$.}
    \label{fast}
    \end{centering}
\end{figure}

This modification results in an eigenfunction that exhibits a discontinuity at the interface, distinguishing it from the behavior seen in slow transitions, as visible comparing \autoref{fast} and \autoref{slow}.

\begin{figure}[t]
\begin{centering}
\includegraphics[width=0.7\textwidth]{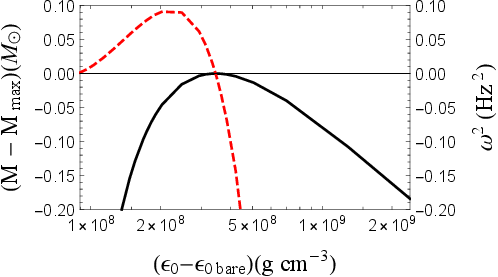}
    \caption{Eigenvalues of the fundamental mode for the slow scenario are depicted by a dashed line, whereas the solid line represents the masses of SDs with a core baryonic content $B_{\text{core}} = 10^{55}$, approaching the Chandrasekhar mass $M_{\mathrm {max}}\approx 0.996 M_\odot$. These are plotted as functions of the central energy density, $\epsilon_0$ shifted by a constant factor $\epsilon_{0\,\, \mathrm{bare}}$ which is a pure QS sharing the same $B_{\text{core}}$. The point where $\omega^2$ becomes zero aligns with the maximum mass, beyond which $\omega^2$ turns negative, indicating instability at higher densities.}
    \label{omegasoslow}
    \end{centering}
\end{figure} 

\begin{figure}[t]
\begin{centering}
\includegraphics[width=0.7\textwidth]{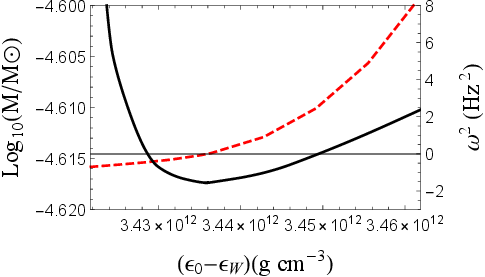}
    \caption{Eigenvalues of the fundamental mode in the fast case (dashed) and masses of SDs having $\varepsilon _t = \varepsilon_\mathrm{drip}$, close to the minimum (solid), plotted as functions of the difference between the central energy density, $\epsilon_0$ and the Witten one $\epsilon_\mathrm{W}$. The zero of $\omega^2$ coincides with the minimum mass. This result is consistent with \cite{Alford:2017vca}}
    \label{omegafast}
    \end{centering}
\end{figure}

The reason behind the apparent inconsistency between the findings of \citet{Glendenning:1994sp,Glendenning:1994zb} and those of \citet{Alford:2017vca} is now evident. In the work of \citet{Alford:2017vca}, they employed an EoS similar to the one discussed in  \autoref{eq:piecewiseEOS}. However, they introduced a smoothing mechanism that eliminated the sharp discontinuity between the two phases, and notably, allowed for an instantaneous transformation from one phase to the other for whatever oscillation timescale. The smoothed EoS used in \cite{Alford:2017vca} can be written as:

\begin{align}
\varepsilon(P)&= \left[1-\mathrm{tanh}\left( (P-P_\mathrm{crit})/\delta P\right)\varepsilon_\mathrm{BPS}(P) \right]/2 \nonumber\\&+
     \left[1+\mathrm{tanh}\left( (P-P_\mathrm{crit})/\delta P\right)\varepsilon_\mathrm{quark}(P) \right]/2
\end{align}

Here, $\delta P$ represents the transition width. This approach is analogous to the rapid transition case discussed here since they are implicitly allowing for a mixed phase. While in \cite{Alford:2017vca}, the eigenfunction doesn't exhibit a discontinuity at the interface, it experiences a very rapid increase in its value. The magnitude of this increase is entirely equivalent to the magnitude of the discontinuity we obtain, which is illustrated as example in  \autoref{fast}.

In contrast, \citet{Glendenning:1994sp,Glendenning:1994zb} did not provide a detailed discussion on the boundary conditions at the interface. However, it is likely that in their work, the eigenfunction was assumed to be continuous, corresponding to the situation described in our "slow" transition scenario.

The distinction between slow and rapid transitions introduced by \citet{Thorne} is based on the observation that the consistency between stability analyses, based on solutions of the TOV equation and those relying on the radial oscillation equation,  is connected to the use of an adiabatic index derived from the EoS employed in the static analysis. They actually coincide in the case of rapid transitions. In contrast, in slow transitions, it is generally challenging to calculate the adiabatic index, primarily due to the need to account for imbalances introduced by perturbations in the computation of the slow adiabatic index \citep{Lindblom:2001hd,Drago:2003wg}.

In our case the conversion between hadrons and quarks is confined to a two-dimensional surface, rather than an extended volume. This simplifies the modification of both the adiabatic index since it corresponds to adapting the interface conditions \citep{Pereira:2017rmp}, and the EoS. In the context of the slow case, this means keeping the quark content in a frozen state \citep{Vartanyan:2009zza,Vartanyan:2012zz}. This dual modification allows us to establish a correspondence between static and dynamic analyses in the slow case, a relationship visually demonstrated in \autoref{omegasoslow}. It's worth noting that this correspondence was already established by \citet{Alford:2017vca} in the context of the rapid transition case and we reproduced the behavior near the minimum in \autoref{omegafast}.

\section{SD collapse}
In a binary system where a WD orbits a main sequence star, mass transfer occurs as the WD accretes material from its companion. This typical scenario culminates in a type Ia supernova (SN) event. However, a different outcome known as accretion-induced collapse (AIC) is theoretically possible. It's important to note that while the concept of AIC has been explored, actual observations of such events are notably absent \citep{Wang_2020}. This absence can be attributed to the substantial difference in timescales between the collapse process and the nuclear reactions responsible for igniting the WD's deflagration.

The presence of a core composed of strange quark matter in SDs plays an important role when the object undergoes significant perturbations, as in the initial stages of a type Ia SN event. In particular, if the quark matter core is large enough, it can potentially facilitate the collapse of the object instead of following the typical path leading to a deflagration. The difficulties of achieving and AIC arises from the fact the nuclear reactions occur when the star is near the Chandrasekhar limit. This phase, characterized by marginal mechanical stability ($\omega^2 \simeq 0$), leads to the star's disruption before AIC can take place \citep{Canal:1990dz}.

The mechanical instability of SDs is strictly related to the rapid conversion of hadrons into quarks. This process is the crucial mechanism that allows the star to undergo a collapse by becoming mechanically unstable.
\begin{figure}[t]
\begin{centering}
    \includegraphics[width=\textwidth]{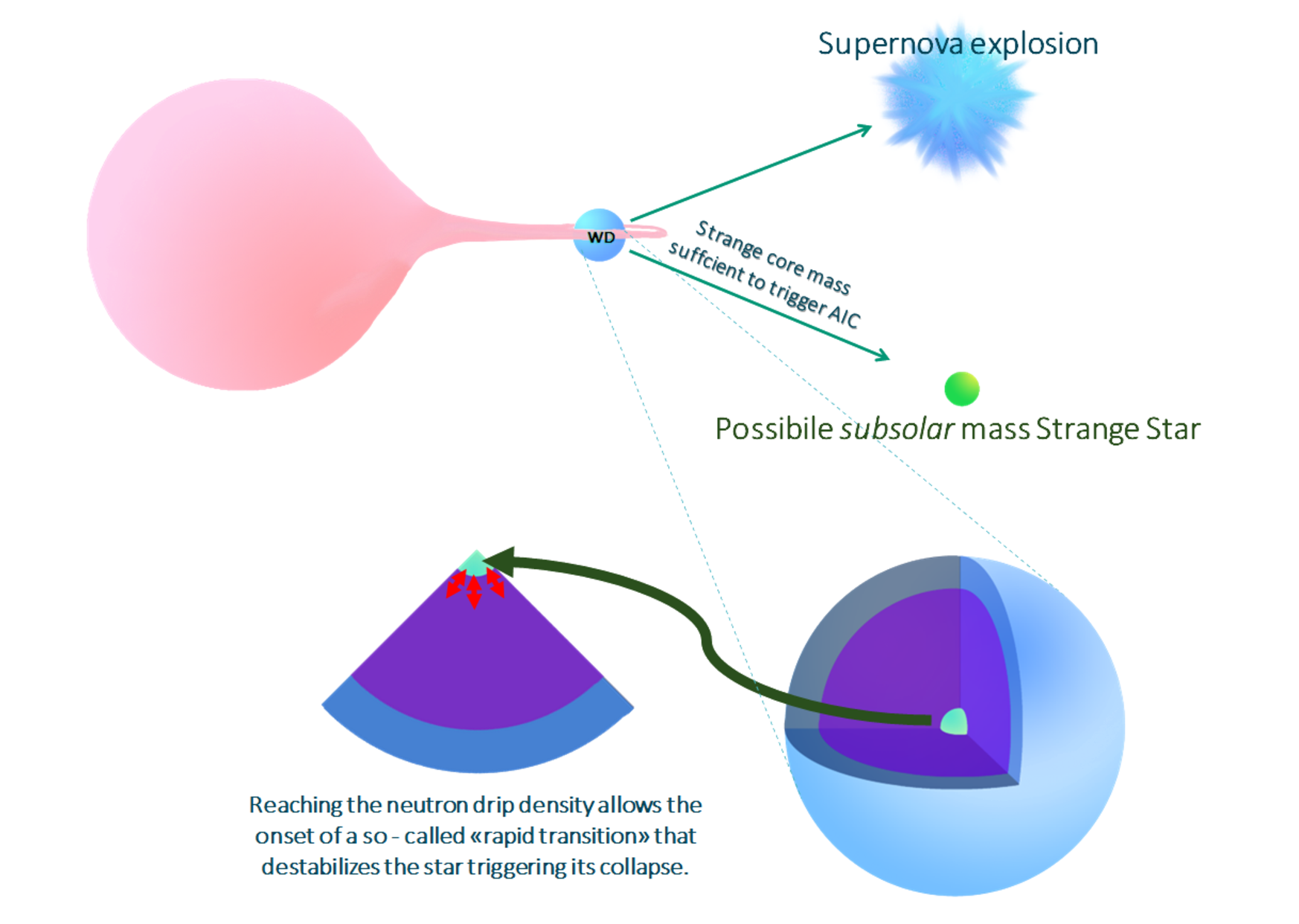}
    \caption{Illustration of the AIC mechanism for a system SD-main sequence star.}
    \label{illustration}
    \end{centering}
\end{figure}

As long as $\varepsilon_\mathrm{t}\ll\varepsilon_\mathrm{drip}$, the object remains mechanically stable. However, if a fluctuation leads to the generation of matter at densities greater than $\varepsilon_\mathrm{drip}$ in a small volume close to the core or if free neutrons are produced (which can fall inside the quark core), the system becomes unstable. To gauge this instability, we calculate the fundamental eigenvalue of a star at the Chandrasekhar limit in the case in which $B_\mathrm{core}$ remains constant. In the case of a slow transition, $\omega^2=0$. Conversely, in the case of a rapid transition, for the same point in the MR diagram, $\omega^2$ becomes significantly negative. It is important to remark that each point at constant $B_\mathrm{core}$ corresponds to a point at constant $\varepsilon_\mathrm{t}$, therefore one can go from a situation in which the transition is physically slow to a situation in which the star internal boundary is in a rapid transition regime and the baryon content of the core is not constant anymore. In any case, it is illogical to apply a slow transition scenario when $\varepsilon_\mathrm{t}$ remains constant, or to employ a fast transition scenario when $B_\mathrm{core}$ is held constant, because $B_\mathrm{core}$ cannot increase in that case and therefore transition must not occur. However, exceptions may arise when the star is in close proximity to the Chandrasekhar mass. In such situations, perturbations could potentially drive a small region of the star, located near the strange core, to exceed the neutron drip density or generate some free neutrons. This, in turn, could trigger the phase transition, therefore going from having a constant $B_\mathrm{core}$ to a situation in which $B_\mathrm{core}$ increases because of the neutron flux.
\begin{figure}[ht!]
\begin{centering}
    \includegraphics[width=0.7\textwidth]{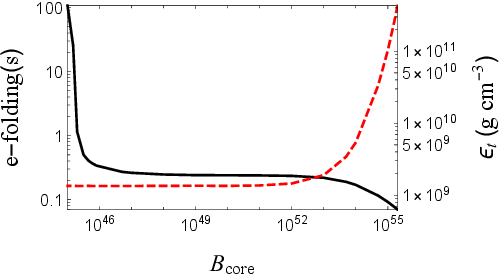}
    \caption{The characteristics of stars at their maximum mass are analyzed in relation to their quark content. The solid black curve represents how the timescale of mechanical instability varies with $B_\mathrm{core}$. Meanwhile, the dashed red line illustrates the density at which the transition occurs, highlighting the relationship between the core's baryonic content and the WD's structural stability.}
    \label{collapse}
    \end{centering}
\end{figure}

\autoref{collapse} displays the e-folding time, defined as $2 \pi/ |\omega |$. It is evident that there exist a threshold that separates a typical SN from an AIC. For $B_\mathrm{core} \gtrsim 10^{46}$, the growth time of the instability falls well below 1 second. This implies that the collapse can happen more rapidly than the full development of a deflagration which is of the order of several seconds \citep{Gamezo:2002nc}. It can be argued that the amount of quark matter that triggers the collapse depends on the EoS. Therefore, we also computed the e-folding time for a different set of parameters for the quark EoS, as presented in \citet{Bombaci:2020vgw} and the results remained consistent.

In the same figure, it is displayed $\varepsilon_\mathrm{t}$, the maximum density reached by the nuclear matter component at the boundary. From the behavior of $\varepsilon_\mathrm{t}$ it is possible to determine when the static structure of a 1 $M_\odot$ SD remains similar to the one of a WD. When $B_\mathrm{core}\gtrsim 10^{52}$ the structure of the star changes and the boundary density $\varepsilon_\mathrm{t}$ deviates from $\sim 10^9$ g/cm$^3$, which is the typical central density of a WD at the Chandrasekhar mass. This suggests that the presence of the quark core does not influence the static properties of the star unless the value of $B_\mathrm{core}$ is large enough to exert a noticeable gravitational pull.

An essential query regarding SDs pertains to the mechanism by which they accumulate the strange quark matter located at their core. The most straightforward explanation lies in the idea that WDs gradually accumulate strangelets over their lifespan. This idea is linked to the possibility that dark matter is made, at least in part, of strangelets \citep{Witten:1984rs}. In a few papers it has been shown that this scenario is compatible with the most recent data from cosmology and astrophysics \citep{Burdin:2014xma,Jacobs:2014yca,SinghSidhu:2019tbr,diclemente2024strange}. In \citet{DiClemente:2022wqp} we have shown that an astrophysical path leading to the formation of subsolar compact stars in electron-capture supernovae can be based on the hypothesis of dark matter made of strangelets. The existence of subsolar mass compact objects has been suggested in \citet{doroshenko2022} and it cannot be explained by solely considering standard equations of state \citep{Suwa:2018uni}.

Another mechanism to produce subsolar-mass compact objects is based on AIC. If AIC takes place in a SD instead of a WD, a SQS is produced instead of a NS. Since this phenomenon is very energetic and the collapsed object is more bound than a NSs, the final object can be a subsolar mass km-sized compact star \citep{DiClemente:2022ktz,DiSalvo:2018mua}. 

\section{Other signatures}
It has been suggested in \citet{Perot:2023gid} a novel approach to distinguish SDs from WDs by utilizing gravitational-wave observations, specifically by measuring the tidal deformability.

When comparing SDs to WDs, a notable feature is the significant reduction in the tidal deformability coefficient which can reach up to 50$\%$ for an SD with a mass of 0.6 $M_\odot$. This difference in tidal parameters could be measured by upcoming space-based gravitational-wave detectors such as the Laser Interferometer Space Antenna (LISA). 

In \citet{Perot:2023gid}, the authors do not employ the BPS EoS to describe WD structure. Instead, they utilize a more refined EoS that accounts for the precise atomic species and their correct balance within the WDs. Additionally, pycnonuclear reactions are taken into account, which usually have their onset at a density of $10^{10}$ g/cm$^3$ for the carbon layer, so at density of interest for SDs. Nevertheless, their analysis indicates that the impact of these reactions is minimal on the SD structure. 

\citet{Perot:2023efc} on the other hand, focus on the effect of  having crystalline color superconductor \citep{Anglani_2014} as the phase for the strange quark matter core. The effect of the large rigidity of the elastic core on the tidal deformability should be relevant (which has a shear modulus that is 2-3 times larger than the one of the hadronic envelope) but it is totally canceled because of the presence of the surrounding hadronic layers. 
Nevertheless, the reduction of the tidal deformability with respect to the one of a WD is still relevant just because of the presence of the strange matter core.

\subsection{Possible SD observations}\label{subsec:kurban}
\begin{figure}[H]
\begin{centering}
    \includegraphics[width=0.7\textwidth]{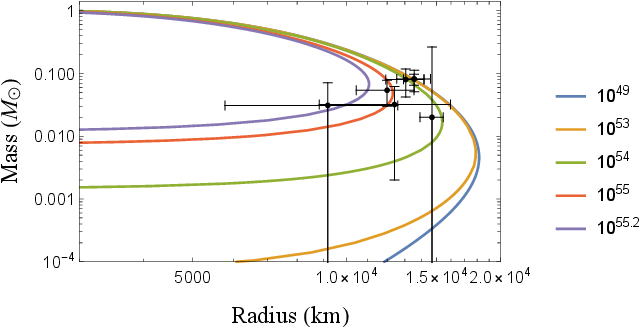}
    \caption{MR relation for constant $B_\mathrm{core}$ SDs with data from \citet{Kurban:2020xtb} analysis.}
    \label{puntikurban}
    \end{centering}
\end{figure}
It is worth noting that in a study by \citet{Kurban:2020xtb}, seven potential SD candidates were identified. These candidates exhibit a mass range spanning from approximately 0.02 up to 0.12 $M_\odot$ and have relatively consistent radii, falling within the approximate range of 9000 to 15000 km. To illustrate the incompatibility with WDs models of those candidates, the paper presents MR relationships for WDs using EoSs of pure magnesium (Mg) and helium (He) stars, in addition to the BPS EoS. By employing an EoS for SDs similar to that in \citet{Alford:2017vca}, it is shown how these objects, due to their compactness, can be good SD candidates. Moreover, this claim is supported from the fact that the BPS EoS serves as an upper limit for compactness in WDs.

To identify potential SD candidates, the authors analyzed WDs listed in the Montreal White Dwarf Database\footnote{\url{http://www.montrealwhitedwarfdatabase.org/tables-and-charts.html}}. They employed analyses based on spectroscopic data and Gaia observations \citep{Blouin:2019}. The data points with their error bars corresponding to the objects identified as SD candidates are visible in \autoref{puntikurban}. 

Recent and more precise measurements of masses and radii of these WDs by the satellite GAIA \citep{gaia3} indicate that some of these objects could be compatible with a normal WD scenario.

\section{Conclusions}

The existence of strange dwarfs is an intriguing possibility in astrophysics since it is strictly connected to cosmology and in particular to the possibility that dark matter is composed by nuggets of strangelets \citep{Witten:1984rs,diclemente2024strange}. One of the most profound astrophysical consequences of the existence of strange dwarfs is the potential formation of subsolar SQSs \citep{DiClemente:2022ktz}. This formation process could occur through an accretion-induced collapse mechanism. However, the feasibility and the specific mechanisms at the basis of this process require further investigation, potentially through detailed studies and simulations in the future.

\vspace{6pt} 

\conflictsofinterest{The authors declare no conflict of interest.}

\abbreviations{Abbreviations}{
The following abbreviations are used in this manuscript:\\

\noindent 
\begin{tabular}{@{}ll}
NS & Neutron Star\\
SQS & Strange Quark Star\\
WD & White dwarf\\
SD & Strange dwarf\\
MR & mass-radius\\
EoS & Equation of state\\
TOV & Tolman-Oppenheimer-Volkoff\\
BPS & Baym-Pethick-Sutherland\\
SN & Supernova\\
AIC & Accretion-induced collapse\\

\end{tabular}
}

\reftitle{References}
\bibliography{references}

\end{document}